*Article*

# Inhomogeneous Broadening of the Exciton Band in Optical Absorption Spectra of InP/ZnS Nanocrystals

**Sergey S. Savchenko[1] and Ilya A. Weinstein[1,2] ***

[1] NANOTECH Centre, Ural Federal University, 19 Mira str., Ekaterinburg 620002, Russia; s.s.savchenko@urfu.ru (S.S.S.)
[2] Institute of Metallurgy, Ural Branch of the Russian Academy of Sciences, 101 Amundsena str., Ekaterinburg 620016, Russia;

* Correspondence: i.a.weinstein@urfu.ru; Tel.: +7-343-375-9374

**Abstract:** In this work, we have simulated the processes of broadening the first exciton band in optical absorption spectra (OA) for InP/ZnS ensembles of colloidal quantum dots (QDs). A phenomenological model has been proposed that takes into account the effects of the exciton-phonon interaction and allows one to analyze the influence of the static and dynamic types of atomic disorder on the temperature changes in the spectral characteristics in question. To vary the degree of static disorder in the model system, we have used a parameter $\delta$, which characterizes the QD dispersion in size over the ensemble. Also, we have calculated the temperature shifts of the maxima and changes in the half-width for the exciton peaks in single nanocrystals ($\delta = 0$), as well as for the integrated OA bands in the QD ensembles with different values of $\delta = 0.6 - 17\%$. The simulation results and the OA spectra data measured for InP/ZnS nanocrystals of 2.1 nm ($\delta = 11.1\%$) and 2.3 nm ($\delta = 17.3\%$) are in good mutual agreement in the temperature range of 6.5 K – RT. It has been shown that the contribution of static disorder to the observed inhomogeneous broadening of the OA bands for the QDs at room temperature exceeds 90%. The computational experiments performed indicate that the temperature shift of the maximum for the integrated OA band coincides with that for the exciton peak in a single nanocrystal. In this case, a reliable estimate of the parameters of the fundamental exciton-phonon interaction can be made. Simultaneously, the values of the specified parameters, calculated from the temperature broadening of the OA spectra, can be significantly different from the true ones due to the effects of static atomic disorder in real QD ensembles.

## 1. Introduction

Semiconductor nanocrystals provide an interesting opportunity for observing the evolution of size-dependent material properties [1-3]. This phenomenon is caused by the quantum confinement of elementary excitations in such small objects. In this context, semiconductor nanocrystals have a size-tunable optical gap, which makes them promising candidates in all applications related to the absorption and emission of light [4-9]. Along with a high luminescence efficiency attainable in quantum dots (QDs), the key parameter is also the half-width of the optical bands, which largely determines the possibility of their use.

In spite of conducting intensive investigations in the field of the synthesis of non-toxic InP-based nanocrystals with high optical characteristics, to interpret their spectra still remains a non-trivial problem [10–14]. In general, in zero-dimensional structures, inhomogeneities and effects inherent to molecular systems significantly contribute to the band broadening of discrete exciton transitions relative to a homogeneous value of the half-width, complicate and overlap the spectra [15,16]. To inhomogeneities specified above, the size-, shape-, stoichiometry-, imperfection-, local atomic environment-, charge state- etc. distributions of nanocrystals over an ensemble are referred [15–21]. These factors affect the transition energy in individual nanocrystals in the ensemble, ultimately causing an inhomogeneous broadening of the experimental optical absorption and luminescence bands. Thus, deep insight into the relationship between the structural characteristics and the optical properties of such systems requires running a comprehensive analysis of homogeneous and inhomogeneous contributions to the broadening.

Earlier, we studied the optical absorption (OA) of water-soluble InP/ZnS core/shell colloidal QDs with an average particle size of 2.1 (QD-1) and 2.3 (QD-2) nm (FSUE "The Research Institute of Applied Acoustics", Dubna). The spectra measurements were taken in the temperature range of 6.5–296 K using a Shimadzu UV-2450 spectrophotometer that operates under the traditional double-beam scheme of measuring the optical density $D = lg(I_0/I)$, where $I_0$ и $I$ are the intensities of the optical radiation incident on and passing through the sample, respectively. During the investigations, a Janis CCS-100/204N closed-loop helium cryostat was used to vary the temperature. A detailed description of the experimental procedure and the results obtained are presented in [22, 23]. Analysis of the behavior of the first exciton absorption band showed that its shift parameters correspond to the bulk and low-dimensional modifications of indium phosphide as the temperature decreases. In this case, the half-width $H$ remains unchanged. This paper presents a simulation of the behavior of the exciton absorption band of InP/ZnS nanocrystals over the experimentally studied temperature range. The model proposed allows for the analysis of the influence of static disorder in the ensemble on the magnitude of the observed broadening of the first exciton absorption band.

## 2. Half-Width of the InP/ZnS Exciton Absorption Band

In Figures 1a and 1b, solid lines represent the experimental dependencies of the optical density on the photon energy for QD-1 at various temperatures in the range under study. The observed features in the above dependencies within the energy range of 2.25–3 eV correspond to the first exciton absorption band. As can be seen, with changing the temperature from 296 to 6.5 K, the first exciton absorption band of the nanocrystal ensemble is shifted towards higher energies. Furthermore, as the temperature lowers, an increase in the visible maximum $D_{max}$ of the optical density is observed (see inset in Figure 1a), and the band shape remains unchanged. The estimation of the $H_{vis}$ and $H_G$ band half-widths was made in two ways: visually by the energy position of $D_{max}$ and by using the Gaussian approximation of the low-energy part of the spectrum, respectively.

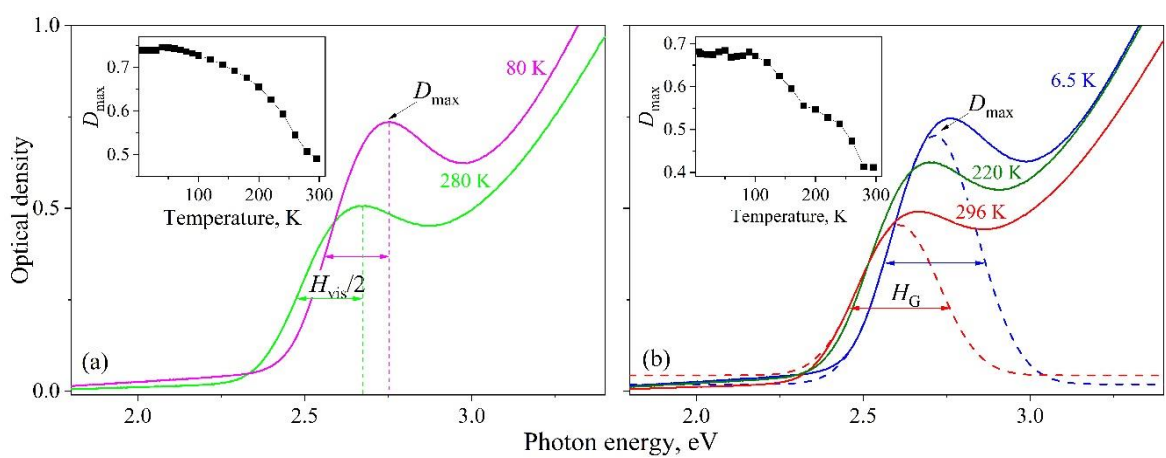

**Figure 1.** Temperature behavior of the first exciton absorption band in InP/ZnS quantum dots. The dashed line shows the approximation by the Gaussian band. The insets show the changes in optical density at the maximum of the exciton band. The experimental data shown were obtained by us in [22].

The $H_{vis}$ value was calculated over the low-energy part of the spectrum. The width at half-height was graphically determined and multiplied by 2 conditional upon that the band is thought to be symmetric (See Figure 1a). The $H_{vis}(T)$ calculated dependence denoted by blue solid squares is shown in Figure 2. It can be seen that the value equal to $H_{vis}$ = 390 ± 10 meV remains constant within the limits of the error above over the entire temperature range under consideration.

The absorption spectrum of QDs is known to exhibit a complex structure of quantized energy levels of electrons and holes. As a rule, exciton bands observed in experiments make themselves felt against the background of contributions from scattering and/or other hidden spectral bands caused by higher-energy transitions [8,24]. Taking these contributions into account, we approximated the low-energy wing of the exciton band by a Gaussian function to quantitatively estimate $H$ (see Figure 1b). The centers of the bands were set at energies $E_1$ obtained using data from second-order derivative spectrophotometry [22,23]. The inset in Figure 1b shows the change in the optical density $D_G$ corresponding to the maximum of the approximating Gaussian. It can be noticed that the temperature dependencies in the insets of Figures 1a and 1b are in qualitative agreement with each other. The $H_G$ values derived from the described analysis of experimental curves are presented in Figure 2; they are denoted by black solid circles. As is clear from the figure, for the QD-1 the $H_G$ = 290 ± 20 meV half-width similar to $H_{vis}(T)$ does not change within the experimental error over the entire temperature range. In this case, $H_G < H_{vis}$ and it corresponds to the half-width of the luminescence band of 250 meV for the QD-1 [22,25]. The same behavior is observed for the QD-2 exciton absorption band with $H_G$ = 370 ± 30 meV [23].

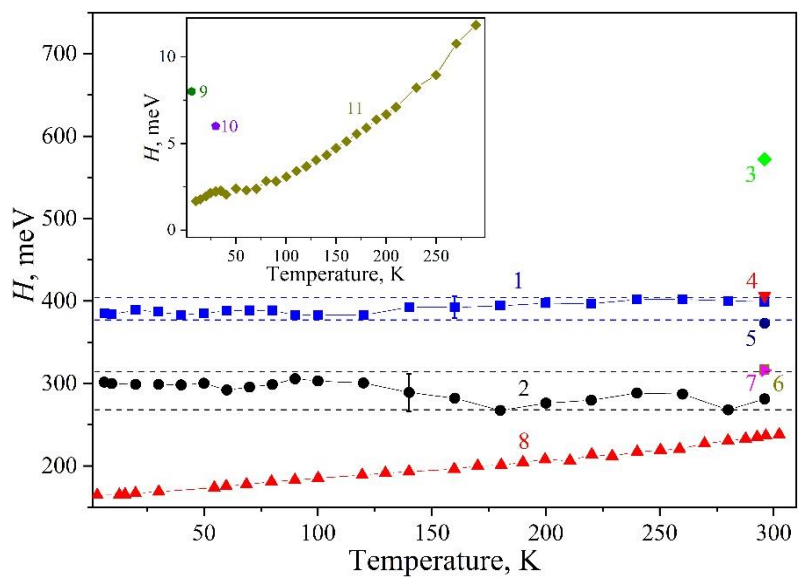

**Figure 2.** The values of $H$ for various InP modifications:
1 - $H_{vis}(T)$ [22],   2 - $H_G(T)$ [22],   3 – InP/ ZnS [6],   4 – InP/ZnS [10],
5 - 3.2 nm InP/ZnS [5],   6 - 2.3 nm InP/ ZnS [27],   7 – InP/ZnS [4],   8 - PL $H(T)$ InP/ZnS [31].
Inset:
9 - InP single crystal [29],   10 - InP nanowires [28],   11 - InP single crystal [30].

Earlier, the temperature-independent exciton band half-width was observed in the OA spectra of CdSSe quantum dots with an average size of 2.3 nm in a glassy matrix [26]. The values of $H$ = 154 meV for the first exciton band that corresponds to the lowest electron-hole pair transition remained almost unchanged during cooling from 300 to 20 K. As the authors point out in their paper, this is due to the dominance of temperature-independent inhomogeneous broadening, mainly related to the size distribution of nanocrystals, $f(a)$.

To calculate the visual half-width values for the first exciton absorption band, we resorted to independent data for InP/ZnS quantum dots of various sizes [4–6,10,27], nanowires [28], and bulk InP single crystals [29,30]. In Figure 2, the different symbols refer to our findings secured. As can be inferred from the figure, the value of $H_{vis}$ at room temperature (RT) (see dependence 1 in Figure 2) is consistent with independent literature data on InP/ZnS core/shell quantum dots (see symbols 3–7 in Figure 2, the average size of nanocrystals, if known, is indicated in the caption). In bulk InP single crystals, the corresponding characteristic varies from 2 to 12 meV in the range of 5–300 K [30], as shown in the inset of Figure 2. Thus, the half-width of the exciton band at 6.5 K in InP/ZnS quantum dots of different sizes and from different manufacturers exceeds more than 100 times the similar characteristic for bulk InP single crystals.

The large value of the exciton absorption band half-width in InP/ZnS quantum dots and its temperature-independence evidences the inhomogeneous nature of the broadening of the optical spectra in the nanocrystals. In general, inhomogeneous broadening in the absorption and emission spectra is typical of nanocrystal ensembles and is a consequence of the nanocrystal distribution in size, shape, stoichiometry, concentration of defects, their charge state, local environment, etc. For a quantitative description of this effect, a model was proposed that reproduces the experimentally observed behavior of the exciton band and allows one to estimate the influence of inhomogeneities in an ensemble of nanocrystals on the magnitude of the optical spectrum broadening.

## 3. Model

### 3.1. Static and Dynamic Disorder in Ensemble

The half-width $h$ of an exciton OA peak of a single nanocrystal at any temperature can be represented as follows:

$$h(T) = h_0 + \Delta_h(T). \tag{1}$$

Here, the first summand reflects the natural line width at zero temperature, and the second one takes into account the effects initiated its temperature broadening. When a system of identical nanocrystals with equal energies $e_1$ of the first exciton transition is considered, the corresponding absorption peak is affected by homogeneous broadening as the temperature changes. Within the exciton-phonon interaction, the corresponding contribution is given by [27, 31, 32]:

$$\Delta_h(T) = \sigma T + A_{LO}\left[\exp(\hbar\omega_{LO}/kT) - 1\right]^{-1}, \tag{2}$$

where $\sigma$ is the excitonic-acoustic phonon coupling coefficient, eV/K; the quantity $A_{LO}$ controls the strength of interaction between excitons and longitudinal optical LO-phonons with the energy $\hbar\omega_{LO}$. The quantity $\Delta_h$ considered herein quantitatively characterizes the dynamic disorder that provides a temperature-dependent contribution to the broadening of energy levels due to lattice vibrations [22].

In real systems, the energy $e_1$ differs for different nanocrystals in an ensemble due to the scatter in the characteristics of QDs [33]. Consequently, the half-width $H$ of the exciton absorption band for the ensemble turns out to be larger than $h$ for individual nanocrystals even at zero temperature. This is because of its forming from a set of various closely spaced, overlapping single-nanocrystal peaks. In the given case, the optical absorption band is inhomogeneously broadened, and the $\Delta_I$ quantity characterizes the static disorder that provides a temperature-independent contribution to the broadening of energy levels [34,35]. The static disorder is caused by the distribution $f$ in terms of QD parameters in the ensemble under study. With increasing temperature, the peaks of individual nanocrystals become wider in accordance with (2), which in turn affects the $H$ exciton band of the ensemble as a whole. This fact leads to a homogeneous contribution of $\Delta_T(f,T)$ to the band broadening. The temperature evolution of the half-width $H$ of the optical absorption band of the nanocrystal ensemble with a certain dispersion $f$ in parameters can be represented as follows:

$$H(f,T) = h_0 + \Delta_I(f) + \Delta_T(f,T). \tag{3}$$

The broadening is governed by the influence of both the static and dynamic types of atomic disorder. In what follows, we will call it temperature broadening.

*3.2. Single-Nanocrystal Exciton Peak Behaviour*

To simulate the temperature behavior of an exciton band of an ensemble of nanocrystals, it is necessary to know the change in the parameters of their individual components. For this, the temperature dependencies should be pre-assigned for the shift of the maximum of individual peaks, their broadening and area changes. The individual spectral components for an ensemble of QDs are assumed to behave identically, depending on temperature.

The shift of the $e_1$ centers of individual Gaussian peaks was modeled in accordance with the Fan expression [22,36,37]:

$$e_1(T) = e_1(0) - A_F \left[\exp(\hbar\omega/kT) - 1\right]^{-1}, \qquad (4)$$

where, $e_1(0)$ is the energy of the first exciton transition at 0 K in a nanocrystal of size *a*, eV; $A_F$ is the Fan parameter depending on the microscopic properties of the material, eV; the expression in parentheses is the Bose-Einstein factor for phonons with an average energy $\hbar\omega$; *k* is the Boltzmann constant, eV/K. As we previously showed in [38], when examining the shift of the first exciton absorption band of the QDs ensemble, the expression describes well the experimental findings for the following parameters specified in Table 1. Figure 3 demonstrates this dependence curve. Thus, the model used and the experiment performed share the same shift magnitude for $e_1$ in the peak for an individual nanocrystal and for $E_1$ in the band for the ensemble as a whole.

**Table 1.** Parameters of temperature evolution for exciton peak in InP/ZnS single nanocrystal.

| Shift | $E_1(0)$, eV | $A_F$, eV | $\hbar\omega$, meV |
|---|---|---|---|
| [22,38] | 2.715 | 0.89 | 15 |
| **Broadening** | $h_0$, meV | $\sigma$, μeV/K | $A_{LO}$, meV | $\hbar\omega_{LO}$, meV |
| [31,39] | 6.3 | 172 | 60 | 40 |

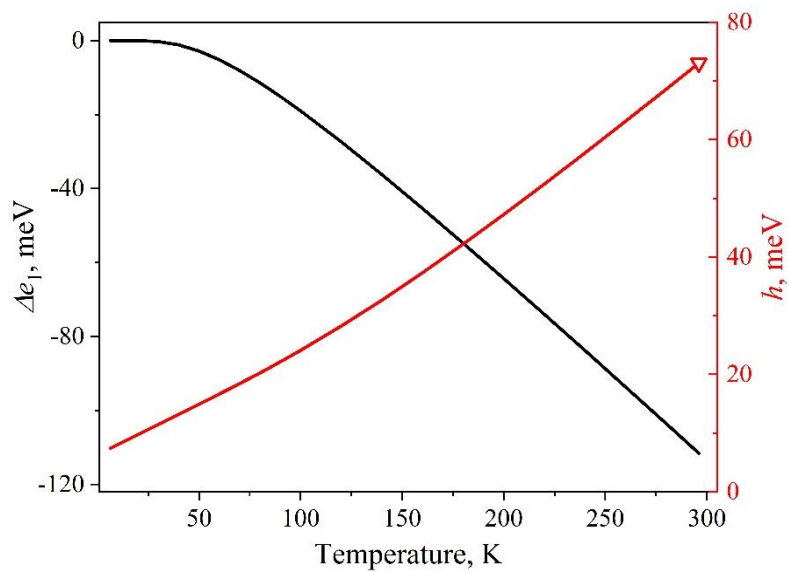

**Figure 3.** Energy shift (black line, calculated by the eq. 4) and broadening (red line, calculated by the eq. 1) of the exciton peak for an individual InP/ZnS nanocrystal. Open triangle – data from [39].

The $h(T)$ dependence was established through the relations (1) and (2). The relation (2) includes the parameter values taken from [31], where they were determined by approximating the $H(T)$ dependence of the luminescence band of an ensemble of InP/ZnS nanocrystals with an average size of 2.1 nm in the temperature range of 2–510 K. The parameters are listed in the Table 1. The value of $h = 73$ meV for an individual InP/ZnS core/shell nanocrystal at room temperature was found in [39] using photon-correlation Fourier spectroscopy (see red open triangle in Figure 3). The close value of $h(300) = 68$ meV was previously obtained for InP nanocrystals when analyzing the size dependence of the bandgap [40]. Figure 3 illustrates the *h*(*T*) dependence calculated for a subset of nanocrystals of the same size in the ensemble. The value of the fundamental half-width $h_0$ in the model amounts to 6.3 meV. The line width at 6.5 K is *h* = 7.4 meV and exceeds the corresponding value for a bulk single crystal (2 meV). The values specified are in complete agreement with the data of [26]. The latter emphasizes that the half-width of optical peaks in quantum dots, measured using the spectral-hole burning technique, is 3 times larger than the bulk analogs.

The temperature-dependent area of the individual components corresponding to single nanocrystals is assumed to change in the same manner as the experimentally observed area *S* of the integral band of the QDs ensemble. The half-width *H* remaining constant over a wide range of temperatures, the *S*(*T*) dependence curve and the behavior of the optical density $D_G(T)$ at the maximum of the exciton band coincide. The *S*(*T*) experimental dependence was obtained as a result of estimating the area of the Gaussians that approximated the first exciton band in the measured optical absorption spectra (see Figure 1). When modeling the temperature behavior of the area of the optical component of a single nanocrystal, an empirical formula was utilized for the relative change in *s*(*T*). This formula describes well the experimentally observed *S*(*T*) dependence:

$$s(T) = \frac{S(T)}{S(0)} = 1 - B_1 \left[\exp(B_2/T) - 1\right]^{-1}, \tag{5}$$

The values of the experimental area normalized at *S*(0) and the corresponding model curve (5) that imitates the *s*(*T*) dependence are presented in Figure 4. The values of the empirical parameters calculated during the approximation are *S*(0) = 0.22 eV, $B_1$ = 0.68 and $B_2$ = 282.92 K. It can be seen that the area of the exciton peak decreases almost twice as the temperature increases in the range studied.

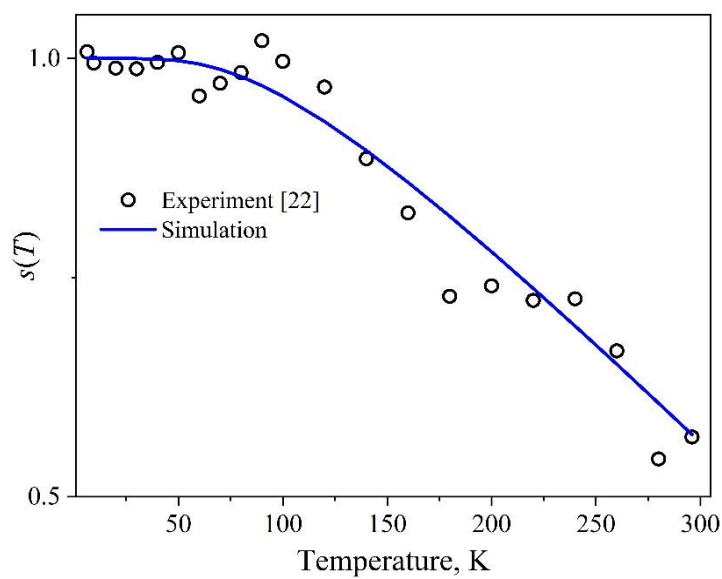

**Figure 4.** Relative change in the exciton peak area calculated by the eq. 5 for InP/ZnS nanocrystals.

*3.3. QDs Size Distribution*

A number of papers regard the scatter of QDs in size *a* as the dominant factor of inhomogeneous broadening [11,16,19,20,33]. Therefore, for definiteness, it is the influence of this factor that will be further associated with the static disorder. The relationship between the size variation in the QD ensemble and the width of the observed exciton absorption band is due, first and foremost, to quantum-size effects. Currently, a number of theories have been proposed to describe the $e_1(a)$ dependence and to take various quantum mechanical approaches and corrections into account [2,3,14,15,41-45]. The present paper applies the well-known expression for estimating the distribution function *f*(*a*) in a QD ensemble through recalculating the exciton transition energy to the size of the corresponding nanocrystal ($e_1 \rightarrow a$). This expression was derived in the effective mass approximation and can be written in the analytical form [15]:

$$e_1(a) = E_g + \pi^2 \left(\frac{a_B}{a}\right)^2 Ry - 1.786 \frac{a_B}{a} Ry - 0.248 Ry, \quad (6)$$

where $E_g$ = 1.42 eV [30] is the band gap of a bulk InP single crystal at 2 K, $a_B$ = 10.1 nm is the Bohr exciton radius, $Ry$ = 5.8 meV is the Rydberg exciton energy for InP, $a$ is the nanocrystal size, nm. For this purpose, the exciton absorption band at 6.5 K was approximated by a Gaussian curve (see Figure 1). In this case, the optical density $D$ at a certain energy is thought to be proportional to the number of nanocrystals of the appropriate size in the ensemble. The normalized distributions $f_1(a)$ and $f_2(a)$ calculated using (6) for the QD-1 and QD-2 ensembles at hand, respectively, are presented in Figure 5 in comparison with experimental data from independent works [46,47]. The average size of the nanocrystals was $\bar{a}_1$ = 2.1 nm and $\bar{a}_2$ = 2.3 nm. To characterize the quality of the size distributions, we evaluated the relative width parameter

$$\delta = \frac{\Delta a}{\bar{a}} 100\%, \quad (7)$$

where $\Delta a$ is the width at half-height for the $f(a)$ dependence. The δ values obtained are shown in Figure 5 in brackets. It can be seen that in the ensembles concerned $\Delta a_2 > \Delta a_1$, $f_1(a)$ and $f_2(a)$ are quite typical of InP/ZnS nanocrystals. It should be noted that such an approach in estimating the size distribution is most appropriate for the absorption band since the latter's position is not affected by the Stokes shift, against the luminescence band. To reduce a temperature influence on the spectral position and half-width of the band, the size distribution needs to be estimated at low temperatures. Then the $\Delta_T(f,T)$ contribution can be neglected.

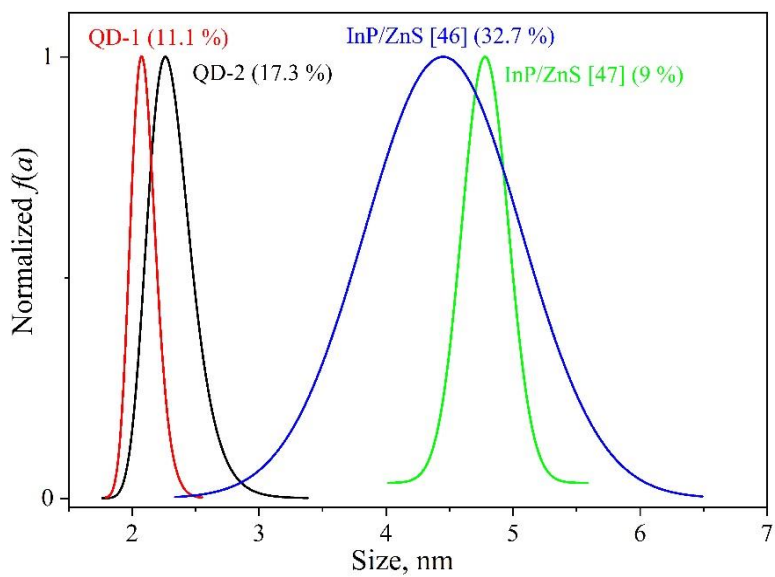

**Figure 5.** Size distributions of the studied InP/ZnS QDs in comparison with the data of independent works. The values of parameter δ calculated by the eq. 7 are given in parentheses.

Varying $H$ of the absorption band underlies modeling the ensembles with different size distributions $f(a)$. In the case of a monodisperse QD ensemble, the absorption peak is formed by a single Gaussian-shaped component with the energy $e_1$ and half-width $h$. The size distribution process leads to dramatically raising the number of such components due to differences in the first exciton transition energy to finally form a Gaussian-shaped band with $E_1$ and half-width $H$.

*3.4. Inhomogeneous and Homogeneous Broadening Contributions*

The influence of the degree of static disorder on the broadening of the first exciton absorption band can be analyzed premised on the model proposed by comparing the temperature and inhomogeneous contributions for the samples with different distributions *f*. The contribution of homogeneous broadening can be quantified as

$$C_H = \frac{\Delta_T(f,T)}{H(f,T)}, \quad (8)$$

The contribution of inhomogeneous broadening, in turn, can be written as

$$C_I = \frac{\Delta_I(f)}{H(f,T)}, \quad (9)$$

Apart from the dimensional criterion δ, we also employed the optical criterion *Q* that reflects the ratio of the homogeneous and inhomogeneous contributions to the band broadening:

$$Q = \frac{C_H}{C_I} = \frac{\Delta_T(f,T)}{\Delta_I(f)}, \quad (10)$$

In this case, it does not matter which factor causes the inhomogeneous broadening since we are dealing with their complex influence on the optical characteristics. For the QD ensembles, this criterion is integral and should monotonously increase with decreasing static disorder in the system. Thus, the criterion can serve as a universal means of comparing the quality of systems with a size effect.

**4. Results and Discussion**

*4.1. Simulation of Experimental InP/ZnS Ensembles*

The simulation of the temperature behavior of the exciton absorption band for a QD ensemble with a given size distribution *f(a)* included three stages. The first one formed the optical absorption (OA) band at zero temperature by a set of Gaussian components. Each of the latter had a half-width $h_0$ and corresponded to a subset of the same size nanocrystals in the ensemble. The second stage determined the behavior of each individual Gaussian as the temperature increases. That is, the position of the maximum of $e_1$, the half-width *h*, and the area *s* were changed according to the functional dependencies established earlier in section *2.2*. At the end, the Gaussian components were summed up at a fixed temperature, and, thus, the model optical exciton absorption band was formed for the QD ensemble with the parameters $E_1$, *H*, and *S*.

The model proposed was intended for describing the temperature behavior of the first exciton absorption band for the experimentally investigated QD-1 ensemble with a size distribution δ = 11.1%. The exciton absorption band with a half-width of 290 meV was approximated by Gaussian components with *h*(6.5) = 7.4 meV, associated with individual nanocrystals in the ensemble. The minimum number of the components to describe the experimental data with high accuracy (Adj. R-square > 0.999) amounted to *N* = 180. Then, taking section 3.2 into account, we simulated the exciton peaks of individual nanocrystals for different experimental temperatures in the range of 6.5–296 K. Figure 6 illustrates the temperature evolution of such an individual spectral component. It can be seen that the maximum shifts by $\Delta e_1$ = 112 meV to the region of lower energies in the temperature range of 6.5–296 K. Simultaneously, the area *s* decreases by 1.8 times, and *h* increases by almost 10 times from 7.4 to 73 meV.

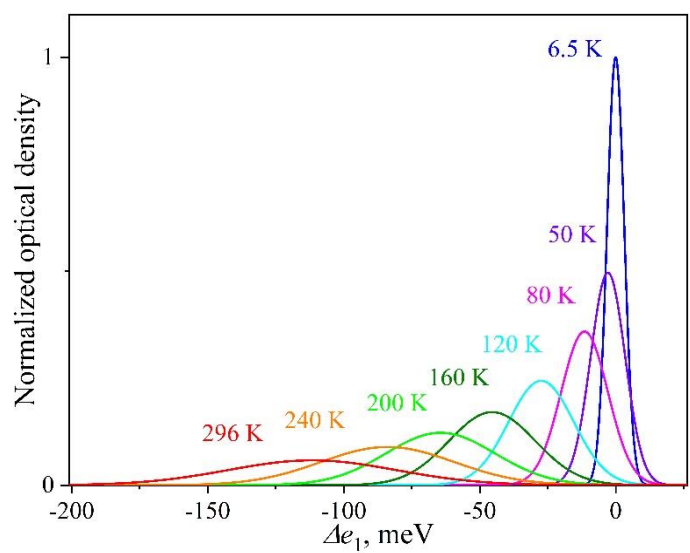

**Figure 6.** Visualization of the temperature behavior of the exciton absorption peak corresponding to individual nanocrystal in the ensemble. For illustrative purposes, the optical density is normalized to the maximum value for a component at 6.5 K.

After summing up the individual spectral components at different temperatures, the temperature behavior of the model integrated band was found to describe well the experimentally observed changes in the parameters of the first exciton absorption band for the InP/ZnS QD ensemble (see Figure 7). The resulting shifts in the position of the model absorption band for the ensemble and the change in its area occur by the same magnitude as in the experiment. It should be noted that the temperature behavior of the integral parameters mentioned above coincides with the behavior of the optical components for single nanocrystals. Thus, good agreement between the experimental and simulation results confirms the validity of the assumptions made earlier. The half-width of the total exciton absorption band for the InP/ZnS QD ensemble in question remains unchanged within the experimental accuracy in the temperature range of 6.5–296 K (see the dependence for $\delta$ = 11.1% in Figure 8 (black open squares)). At the same time, the half-width *h* of the optical peak for each of the nanocrystals in the distribution varies by almost 10 times. The model proposed reproduces properly the temperature behavior of the first exciton absorption band for the InP/ZnS QD ensemble and can be applied for analyzing the influence of the size distribution of nanocrystals due to the contribution of static disorder on the broadening of optical spectra.

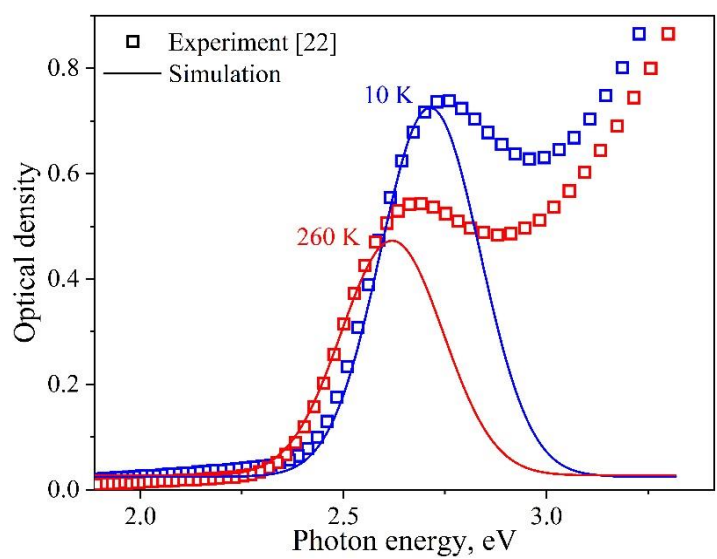

**Figure 7.** Experimental OA spectra [22] (symbols) and simulation of the first exciton absorption band (solid lines) for QD-1 at specified temperatures.

*4.2. Inhomogeneous and Homogeneous Broadening*

To analyze the influence of the degree of static disorder on the homogeneous and inhomogeneous broadening processes in the absorption band, we investigated a number of QD ensembles with different half-widths $H$ of the first exciton band, see Table 2. The value of $N$ indicates the number of Gaussian components used for simulating the corresponding integrated band. In this case, the latter's maximum was at an energy of 2.71 eV, which conforms to the experimental position for the ensemble of the nanocrystals studied at 6.5 K. This means that the computational experiments save the average particle size standing, only the width of the size distribution changes, which is, in turn, reflected in the change in $H$.

Figure 8 presents the results of modeling the temperature change in the half-width of the exciton absorption band for ensembles of nanocrystals with different degrees of static disorder, i.e. by different distributions of $f(a)$, in the temperature range of 6.5–296 K. It can be seen that, as the width of $f(a)$ for nanocrystals in the ensemble diminishes, the homogeneous contribution to the broadening of the model band is more pronounced. In the case of the band (290 meV) experimentally observed, $C_H$ = 3% and does not exceed the experimental error of 7%. With reducing the degree of static disorder, the contribution of homogeneous broadening rises. The $C_H$, $C_I$ and $Q$ calculation findings according to eqs. (8), (9) and (10) at RT for the model range of the samples are given in Table 2. As is clear from the Table, as δ decreases, the behavior of $H$ tends to be characteristic of the only homogeneously broadened component. Otherwise speaking, $H$ behaves as matching to a monodisperse ensemble. On the other hand, the increasing degree of the static disorder in the ensemble leads to lesser and lesser impact of the homogeneous broadening of the peak of each individual nanocrystal on the temperature broadening of the integrated band.

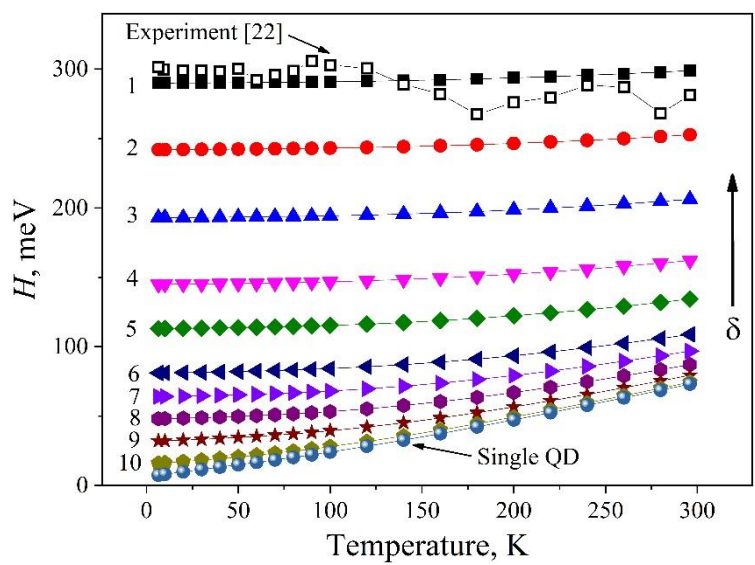

**Figure 8.** Simulated $H(T)$ dependencies of the exciton band of QD ensembles with different size distributions, see Table 2. Open squares denote the experimental dependence for the QD-1 [22].

The simulated regularities also explain the temperature behavior of $H$ of the QD-2 ensemble. For it, the half-width of the optical absorption band is 370 meV, which corresponds to a wider size distribution with δ = 17.3% than for QD-1. The contribution of the $C_H$ to the temperature broadening in such an ensemble is even less than in the case of QD-1.

It should be emphasized that the work [31] reports on a change in the half-width of the luminescence band of an ensemble of InP/ZnS nanocrystals by 0.08 eV in the temperature range of 2–300 K (see curve 8 in Figure 2). This may be due to the transfer of excitation energy over an ensemble of quantum dots from small to larger ones [48]. As a consequence, a certain size subset of nanocrystals participates in the radiative process, which leads to a decrease in the inhomogeneous broadening contribution to the $H$ value of the luminescence band of the ensemble. In this case, the dynamic disorder contribution to the temperature broadening increases.

Table 2. Results of modeling the temperature behavior of exciton absorption band in InP/ZnS QD ensembles with different size distributions (data are shown on Fig. 8).

| Curve | $H$(6.5 K), meV | $N$ | $\delta$, % | $C_I$, % | $C_H$, % | $Q$ |
|---|---|---|---|---|---|---|
| 1 | 290 | 180 | 11.1 | 94.9 | 3.0 | 0.03 |
| 2 | 242 | 150 | 9.2 | 93.3 | 4.2 | 0.05 |
| 3 | 193 | 120 | 7.3 | 90.5 | 6.4 | 0.07 |
| 4 | 145 | 90 | 5.5 | 85.5 | 10.6 | 0.12 |
| 5 | 113 | 70 | 4.3 | 79.4 | 15.9 | 0.20 |
| 6 | 81 | 50 | 3.1 | 68.7 | 25.5 | 0.37 |
| 7 | 64 | 40 | 2.4 | 59.6 | 33.9 | 0.57 |
| 8 | 48 | 30 | 1.8 | 47.9 | 44.9 | 0.94 |
| 9 | 32 | 20 | 1.2 | 32.4 | 59.7 | 1.84 |
| 10 | 16 | 10 | 0.6 | 13.0 | 78.5 | 6.02 |
| single QD | 7.4 | 1 | 0 | 0 | 89.9 | - |

The analysis conducted indicates some limitations that should be considered when analyzing the temperature evolution of the exciton absorption bands in the ensembles of semiconductor nanocrystals. In describing the temperature curves for the shift of the maxima, the values of the extracted fundamental parameters of the exciton-phonon interaction are physically justified. As was shown in the framework of the model proposed, the degree of static disorder in the system does not affect essentially these dependencies. The shift of the integrated band maximum reflects the displacement of the optical component of the peak for an individual nanocrystal. However, when considering the temperature variations in the band half-width for a QDs ensemble, the static disorder plays a significant role. When forming, the integrated absorption band losses the homogeneous broadening of the individual nanocrystal components due to their overlap. The contribution of static disorder distorts the corresponding parameter values of the exciton-phonon interaction, whilst they characterize the ratio of the homogeneous and inhomogeneous broadening processes. It is therefore mandatory to carry out estimates of them carefully. Thus, to make a reliable estimate of the parameters of dynamic disorder, an analysis of the temperature broadening processes in optical spectra needs to run for ensembles with a narrow size distribution of quantum dots.

## 5. Conclusions

A phenomenological model has been proposed that allows one to quantitatively describe the temperature behavior of the first exciton absorption band of an InP/ZnS QD ensemble and to analyze homogeneous and inhomogeneous contributions to the broadening of the band with temperature. Using numerical analysis, the half-width of the exciton absorption band for the InP/ZnS QD ensemble under study has been found to remain unchanged in the temperature range of 6.5–296 K even taking into account the 10-fold homogeneous broadening of spectral components for individual nanocrystals. The result obtained is in agreement with our experimental data as to the behavior of the exciton band of the ensemble of InP/ZnS nanocrystals with a size distribution $\delta$ = 11.1%. The temperature behavior of optical absorption bands for QDs ensembles with different degrees of static disorder due to a variation of $\delta$ has been simulated. The influence of the parameters specified on the model band broadening has been analyzed. It has been shown that the homogeneous contribution to the integrated broadening decreases from 90 to 3% when passing from a monodisperse ensemble of nanocrystals to a distribution with $\delta$ = 11.1%. The experimentally observed half-width $H$ of the exciton band for the InP/ZnS QD ensemble has been established to be unchanged over a wide temperature range due to inhomogeneous broadening processes through a wide size distribution of nanocrystals.

**Author Contributions:** conceptualization, I.W.; software, S.S.; formal analysis, S.S.; investigation, S.S.; resources, I.W.; writing—original draft preparation, S.S. and I.W.; writing—review and editing, S.S. and I.W.; visualization, S.S. and I.W.; supervision, I.W.; funding acquisition, S.S. and I.W.

**Funding:** This research was supported by RFBR according to the research project № 18-32-00664 and Act 211 Government of the Russian Federation, contract no. 02.A03.21.0006. I.W. thank Minobrnauki initiative research project № 16.5186.2017/8.9 for support.

**Conflicts of Interest:** The authors declare no conflict of interest. The funders had no role in the design of the study; in the collection, analyses, or interpretation of data; in the writing of the manuscript, or in the decision to publish the results.